\documentclass[sn-nature]{sn-jnl}

\usepackage{graphicx}%
\usepackage{multirow}%
\usepackage{amsmath,amssymb,amsfonts}%
\usepackage{amsthm}%
\usepackage{mathrsfs}%
\usepackage[title]{appendix}%
\usepackage{xcolor}%
\usepackage{textcomp}%
\usepackage{manyfoot}%
\usepackage{booktabs}%
\usepackage{algorithm}%
\usepackage{algorithmicx}%
\usepackage{algpseudocode}%
\usepackage{soul}%
\usepackage{setspace}
\usepackage{listings}%

\theoremstyle{thmstyleone}%
\theoremstyle{thmstyletwo}%

\theoremstyle{thmstylethree}%

\begin{document}

\title[Article Title]{Polyepitaxial grain matching to study the oxidation of uranium dioxide}

\author*[1]{\fnm{Jacek} \sur{Wasik}}\email{jacek.wasik@bristol.ac.uk}
\author[1]{\fnm{Joseph} \sur{Sutcliffe}}

\author[2]{\fnm{Renaud} \sur{Podor}}

\author[1,3]{\fnm{Jarrod} \sur{Lewis}}

\author[1,3]{\fnm{James Edward} \sur{Darnbrough}}

\author[1]{\fnm{Sophie} \sur{Rennie}}

\author[1]{\fnm{Syed} \sur{Akbar Hussain}}

\author[1]{\fnm{Chris} \sur{Bell}}

\author[1,4]{\fnm{Daniel Alexander} \sur{Chaney}}

\author[1]{\fnm{Gareth} \sur{Griffiths}}

\author[1]{\fnm{Lottie Mae} \sur{Harding}}

\author[1]{\fnm{Florence} \sur{Legg}}

\author[1,4]{\fnm{Eleanor} \sur{Lawrence Bright}}

\author[1]{\fnm{Rebecca} \sur{Nicholls}}

\author[5]{\fnm{Yadukrishnan} \sur{Sasikumar}}

\author[1]{\fnm{Angus} \sur{Siberry}}

\author[1]{\fnm{Philip} \sur{Smith}}

\author*[1]{\fnm{Ross} \sur{Springell}}\email{phrss@bristol.ac.uk}

\affil[1]{\orgdiv{IAC, School of Physics}, \orgname{University of Bristol}, \orgaddress{\street{Tyndall Avenue}, \city{Bristol}, \postcode{BS8 1TL}, \state{Avon}, \country{UK}}}

\affil[2]{\orgname{ICSM}, \orgaddress{\street{Bagnols-sur-Cèze}, \city{Site de Marcoule}, \postcode{30207}, \country{France}}}

\affil[3]{\orgname{Department of Materials}, \orgname{University of Oxford} \orgaddress{\street{21 Banbury Road}, \city{Oxford}, \postcode{OX2 6NN}, \country{UK}}}

\affil[4]{\orgname{European Synchrotron Radiation Facility}, \orgaddress{\street{71 Avenue des Martyrs}, \city{Grenoble}, \postcode{38000}, \country{France}}}

\affil[5]{\orgname{Oak Ridge National Laboratory}, \orgaddress{\street{1 Bethel Valley Road}, \city{Oak Ridge}, \postcode{TN 37831}, \country{US}}}


\abstract{ 

Although the principal physical behaviour of a material is inherently connected to its fundamental crystal structure, the behaviours observed in the real-world are often driven by the microstructure, which for many polycrystalline materials, equates to the size and shape of the constituent crystal grains. Here we highlight a cutting edge synthesis route to the controlled engineering of grain structures in thin films and the simplification of associated 3-dimensional problems to less complex 2D ones. This has been applied to the actinide ceramic, uranium dioxide, to replicate structures typical in nuclear fission fuel pellets, in order to investigate the oxidation and subsequent transformation of cubic UO$_{2}$ to orthorhombic U$_{3}$O$_{8}$. This article shows how this synthesis approach could be utilised to investigate a range of phenomena, affected by grain morphology, and highlights some unusual results in the oxidation behaviour of UO$_{2}$, regarding the phase transition to U$_{3}$O$_{8}$.}

\keywords{Polyepitaxy, Uranium oxides, Oxidation}



\maketitle

\section{Introduction}\label{sec1}

The nature of grains and grain boundaries is crucial to our understanding of the behaviour of polycrystalline materials \cite{clarke1987grain, watanabe2011grain, cantwell2020grain, krause2019review}. The relationship between these defect dense regions and the misaligned crystal grains can be affected by geometries, such as tilt and twist angles, and by impurities, which tend to precipitate and segregate \cite{yajima1966lattice, liu2021grain}. All this complexity can have dramatic consequences on critical properties, such as mechanical strength, electrical and thermal conductivities, and corrosion resistance \cite{yajima1966lattice, liu2021grain, sabioni2000effect, millett2012grain, andersson2015multiscale, lim2016effects}. In this paper, we demonstrate an approach to controlled grain engineering, which reduces a three dimensional problem to two dimensions, yielding new results in an important applied materials system in the nuclear industry.

Uranium dioxide is the most prevalent fission fuel and is used world-wide in over 400 nuclear reactors \citep{bayraktar2023nuclear}. Typical fuel pellets are small cylinders of 1\,cm height and diameter and are placed inside fuel pins several metres long. There are tens of thousands of fuel pins in a reactor core, resulting in more than 10 million fuel pellets in an average, water-moderated reactor \citep{bayraktar2023nuclear}. This means that there are more than 30,000 tonnes of UO$_{2}$ fuel currently in operation. This fuel is most commonly manufactured by sintering powdered UO$_{2}$ at high temperature and pressure to achieve densities approaching 97\% \citep{mceachern1998review}.

\begin{figure}[h!]%
 \centering
 \includegraphics[width=\textwidth]{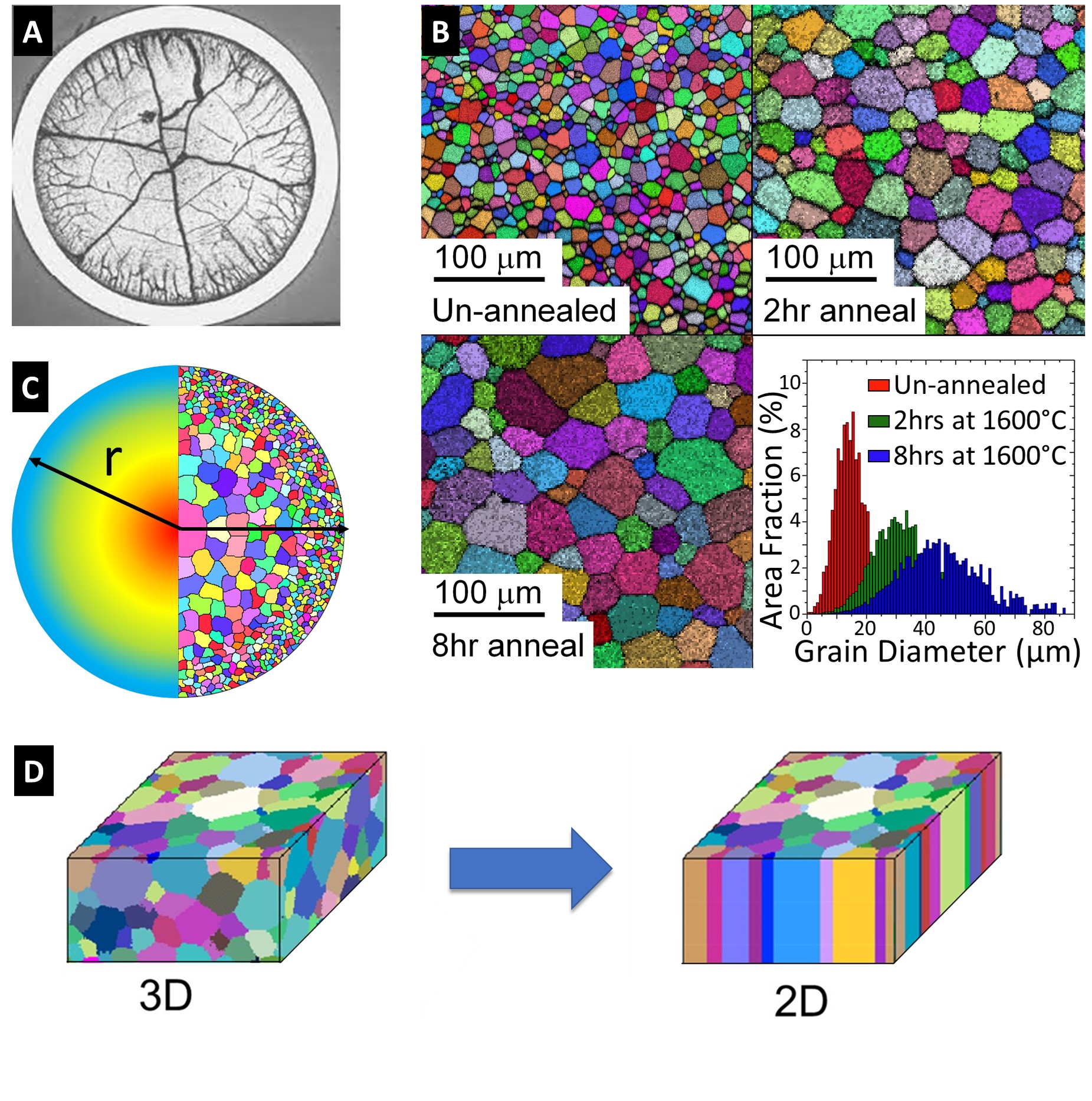}
 \caption{\textbf{Variation of UO\textsubscript{2} grain size within nuclear fuel and replication method.}(A) shows a cross-section of irradiated nuclear fuel \cite{suzuki2008thermal} (B) EBSD data collected from UO\textsubscript{2} thin films samples deposited onto un-annealed, 2h and 8h-annealed polycrystalline YSZ substrates, with an average grain size of 14.2$\pm$0.2 $\mu$m, 31.7$\pm$0.3 $\mu$m and 45.8$\pm$0.4 $\mu$m, respectively, demonstrating the ability to engineer samples with desired grain structure (C) schematic representation of thermal gradient (red $\>1500\,^{\circ}$C $\rightarrow$ blue $\>300\,^{\circ}$C) across the fuel pellet and resulting increase in the uranium dioxide grain size  (D) schematic of the reduction of a three dimensional granular structure into two dimensions by producing nanometre thick columns of micon-size diameter grains.}\label{Fig1}
 \end{figure}

During operation the fission process causes severe structural transformations, driven by the kinetic energy of daughter nuclei and emitted neutrons. There is an enormous amount of heat deposited in the fuel pellets, which is transported away by the coolant in the core, however, the fundamental thermal conductivity of UO$_{2}$ is particularly poor and this results in large thermal gradients across only a few mm. Centreline temperatures can reach $\>1500\,^{\circ}$C with the outer edge of the fuel pellet just a 300$^{\circ}$C \citep{wilson1996nuclear}. These effects cause grain growth in the centre of the pellet, with average grain sizes of about 50\,microns, becoming smaller radially outwards, down to an average size of 10\,microns in the so-called rim-structure \citep{devanathan2010modeling, suzuki2008thermal, mceachern1998review, sonoda2002transmission, spino1996detailed} (see Fig. \ref{Fig1}A and \ref{Fig1}C). The grain structure is particularly important in two crucial stages of the fuel lifetime; 1) affecting the thermal transport and migration of fission products during operation and 2) affecting oxidation and corrosion post-operation during interim and final storage, especially in aqueous environments. \citep{devanathan2010modeling, suzuki2008thermal, mceachern1998review, sonoda2002transmission, spino1996detailed, shoesmith2000fuel, springell2015water}.

The use of thin film deposition techniques to synthesise fission fuel materials is gaining traction as a viable and, in some cases, preferred route for test and experimentation \citep{vallejo2022advances, springell2023review}. The bulk fuel material is complex making it difficult to deconvolve the many contributing factors. Spent nuclear fuel presents even more of a problem as the handling of such highly radioactive material is only possible in a handful of specialist laboratories. A thin film approach to fuel synthesis creates a macroscopic surface with limited thickness, typically tens to hundreds of nanometres, where radioactivity levels are very low; a 100\,nm thick, 1\,cm$^{2}$ sample is 1.5\,Bq. It is then possible to control structural and chemical parameters, such as crystallographic orientation, strain, and stoichiometry \citep{springell2023review}. We present a unique mode of growth that can engineer the grain size of high temperature ceramics, such as UO$_{2}$, without requiring the extremely high temperatures required in bulk materials. This process relies on a 1:1 substrate match, combined with careful heat treatment and surface preparation of substrate materials, before deposition.

The pre-treatment of the substrate material can be used to selectively engineer desired average grain sizes (see Fig. \ref{Fig1}C). The columnar-type growth of these epitaxial grains means that although the lateral grain dimensions mimic those of the underlying polycrystal, the in-plane microstructure is uniform along the growth direction. This approach has distinct advantages for any depth-dependent studies, as a complex 3D problem is effectively reduced to two dimensions (see Fig. \ref{Fig1}D). This approach can be adapted to a wide variety of materials, where grain size-dependent effects are of interest. We have applied this method to uranium dioxide, which is most familiar in a small cylindrical sintered pellet form for nuclear fuel. Here we present the first images of UO$_{2}$ synthesised in this way in order to investigate the grain orientation-dependent oxidation to U$_{3}$O$_{8}$. This process follows the oxidation sequence as: UO\textsubscript{2} $\Rightarrow$ U\textsubscript{4}O\textsubscript{9} $\Rightarrow$ U\textsubscript{3}O\textsubscript{7} $\Rightarrow$ U\textsubscript{3}O\textsubscript{8}, with nucleation and growth kinetics \citep{mceachern1998review, rousseau2006detailed, allen1986formation, allen1987oxidation, allen1995mechanism, geonvold1948oxidation, leinders2016low, idriss2010surface, liu2021grain, lim2016effects, quemard2009origin}.

This is of direct industrial interest in the initial processing and eventual disposal of nuclear fuel. The precise mechanism of UO$_{2}$ oxidation is still not fully understood and there are outstanding questions regarding the transformation of UO$_{2}$ to U$_{3}$O$_{8}$. Here we have discovered some peculiar results, which could shed light on a specific crystallographic route for this phase transition.

\section{Results}\label{sec2}

\subsection{Sample synthesis and grain engineering}

The UO$_{2}$/YSZ system (YSZ is yttria-stabilised zirconia) is an example of a 1:1, or so-called cube on cube matched epitaxial relationship, used to produce single crystal films to study surface and interfacial effects on pristine fuel material \citep{strehle2012characterization, rennie2018role}. The ability to match grains of UO\textsubscript{2} onto YSZ is demonstrated in Fig. \ref{Fig1}B, which  shows electron-backscatter images from the surface of 100\,nm polycrystalline films of stoichiometric UO$_{2}$, where the final deposited surface requires no further mechanical treatment (avoiding possible influence of mechanical polishing and defect formation), typically necessary to achieve high quality Electron Backscatter Diffraction (EBSD) patterns from bulk material. Polycrystalline substrates of YSZ were heat-treated and polished flat before reactive dc magnetron sputtering of UO$_{2}$, which required no further treatment to get the observed patterns. The three samples shown here are chosen to represent three positions across a fuel pellet and have grain orientation distribution akin to that found in bulk material, but with average grain sizes of 14.2$\pm$0.2 $\mu$m, 31.7$\pm$0.3 $\mu$m and 45.8$\pm$0.4 $\mu$m (errors here are represented by the standard error). In a bulk material any measurement of the surface properties that is any deeper than the first few atomic layers will be affected by the grain structure along the surface normal. Our polyepitaxial growth mechanism avoids these issues, creating columnar grains matched along the underlying YSZ grain directions, where the lateral grain structure remains constant as a function of depth (see Fig. \ref{Fig1}D). This provides an ideal scenario to study and engineer grain effects for a wide range of materials properties. Here we focus on the effect of grain size, orientation and grain boundary volume on the oxidation of UO$_{2}$.

\begin{figure}[h]%
\centering
\includegraphics[width=\textwidth]{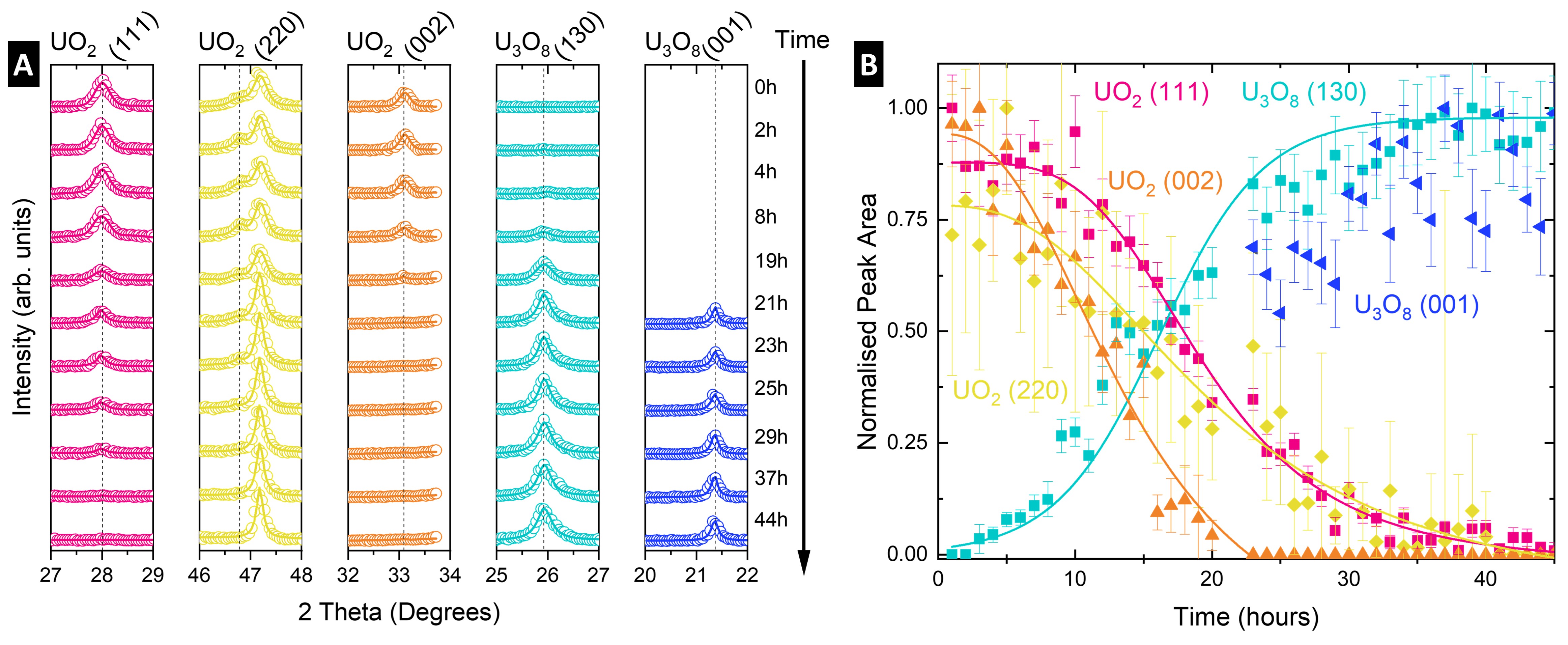}
\caption{\textbf{\textit{In-situ} XRD during oxidation of polyepitaxial UO\textsubscript{2} thin film.} (A) Example longitudinal XRD patterns collected during oxidation polyepitaxial UO\textsubscript{2} thin film at 200 mbar of oxygen at 300\textdegree{}C over 48 hours. The right peak on the UO\textsubscript{2} (220) column is associated with the XRD sample stage. Data is represented by open circles and fit by line. (B) Normalized Bragg's peaks area for 3 main orientations of cubic UO\textsubscript{2} system and orthorombic U\textsubscript{3}O\textsubscript{8} ploted over time. The data shows faster oxidation along (002) orientation, and characteristic nucleation and growth kinetics}\label{Fig2}
\end{figure}

\subsection{X-ray diffraction investigation of oxidation process}

The lattice parameter of the as-grown samples was 5.476(2)\AA{}, similar to values reported for the bulk material \citep{mceachern1998review, barrett1982preparation, swanson1955standard, desgranges2009neutron}. The polycrystalline character of the sample was confirmed by comparing measured integrated intensities with predicted structure factor intensities. Based on the (111) being the strongest intensity 100\%, the measured values for (220) and (002) were 48$\pm{}$3\% and 34$\pm{}$3\%, with expected theoretical values 47\% and 36\% respectively, exhibiting minimal preferential texture. The lattice parameter as a function of x in the UO\textsubscript{2+x} system was observed to have non-Vegard character as reported by Elorrieta et al. \citep{elorrieta2016detailed}. The oxidation of polyepitaxial UO\textsubscript{2} thin films was performed in two stages. In the first stage, the sample was exposed to 200 mbar of O\textsubscript{2} at 150\textdegree{}C for 16 hours. (see supplementary data). The lattice constant first shrinks in the region below x = 0.13, and then slightly expands for 0.13 $<$ x $<$ 0.17 before shrinking again for x above 0.17. The final product at this stage was identified as tetragonal U\textsubscript{3}O\textsubscript{7}, as c is 1.9\% longer than a \citep{smith1982phase, leinders2016assessment, leinders2021charge}. No further oxidation was observed, as expected in this temperature range \citep{rousseau2006detailed, lim2016effects}.

In the second stage of the experiment, the pressure of oxygen was kept constant at 200 mbar, and the temperature was increased to 300\textdegree{}C. The Longitudinal scans were collected every 0.5 hour for 48 hours, and a selection of these are shown in Fig. \ref{Fig2}A. A general decrease in the intensity is observed for (111), (002) and (220) reflection assigned to the UO\textsubscript{2+x} structure as the cubic/tetragonal structure undergoes a phase transition to the orthorhombic structure of U\textsubscript{3}O\textsubscript{8}. This coincides with an increase in the observed intensities of the U\textsubscript{3}O\textsubscript{8} reflections within the XRD pattern. To better illustrate this process, an area under each peak was extracted, normalised to the highest value and plotted against time as shown Fig. \ref{Fig2}B.   

The kinetics of oxidation from UO\textsubscript{2+x} to U\textsubscript{3}O\textsubscript{8} presented in Fig. \ref{Fig2}B. are in good agreement with the sigmoidal nucleation-and-growth mechanism described in literature for this phase transition \citep{mceachern1998review, aronson1957kinetic, hoekstra1961low, rousseau2006detailed, desgranges2011refinement, walker1965oxidation, allen1995mechanism}. The induction period, where the initial rate of oxidation is very low, takes place during first the 10 hours of the oxidation process. After this time, the oxidation rate gradually increases to a maximum, and this linear region \citep{mceachern1998review} is observed from around 13 hours to 21 hours. When the reaction approaches completion, it tails off, finishing after 44 hours with the sample completely oxidised to U\textsubscript{3}O\textsubscript{8}.

Data presented here is in contradiction to what was reported by Allen et al. \citep{allen1986formation, allen1987oxidation}, where the oxidation rate of UO\textsubscript{2} was suggested to be highest for the [111] orientation due to epitaxial relations with U\textsubscript{3}O\textsubscript{8}, followed by [110] and slowest for [001]. The [001] oriented grains within the polyepitaxial sample oxidised quickest, and no UO\textsubscript{2+x} \{002\} reflection was observed after 23 hours. There is no significant difference between kinetics observed for the [111] and [110] oriented grains. This might suggest that there is potentially crystallographic relationship between the [001] orientation of UO\textsubscript{2+x} and orthorhombic U\textsubscript{3}O\textsubscript{8}.

\begin{figure}[h]%
\centering
\includegraphics[width=\textwidth]{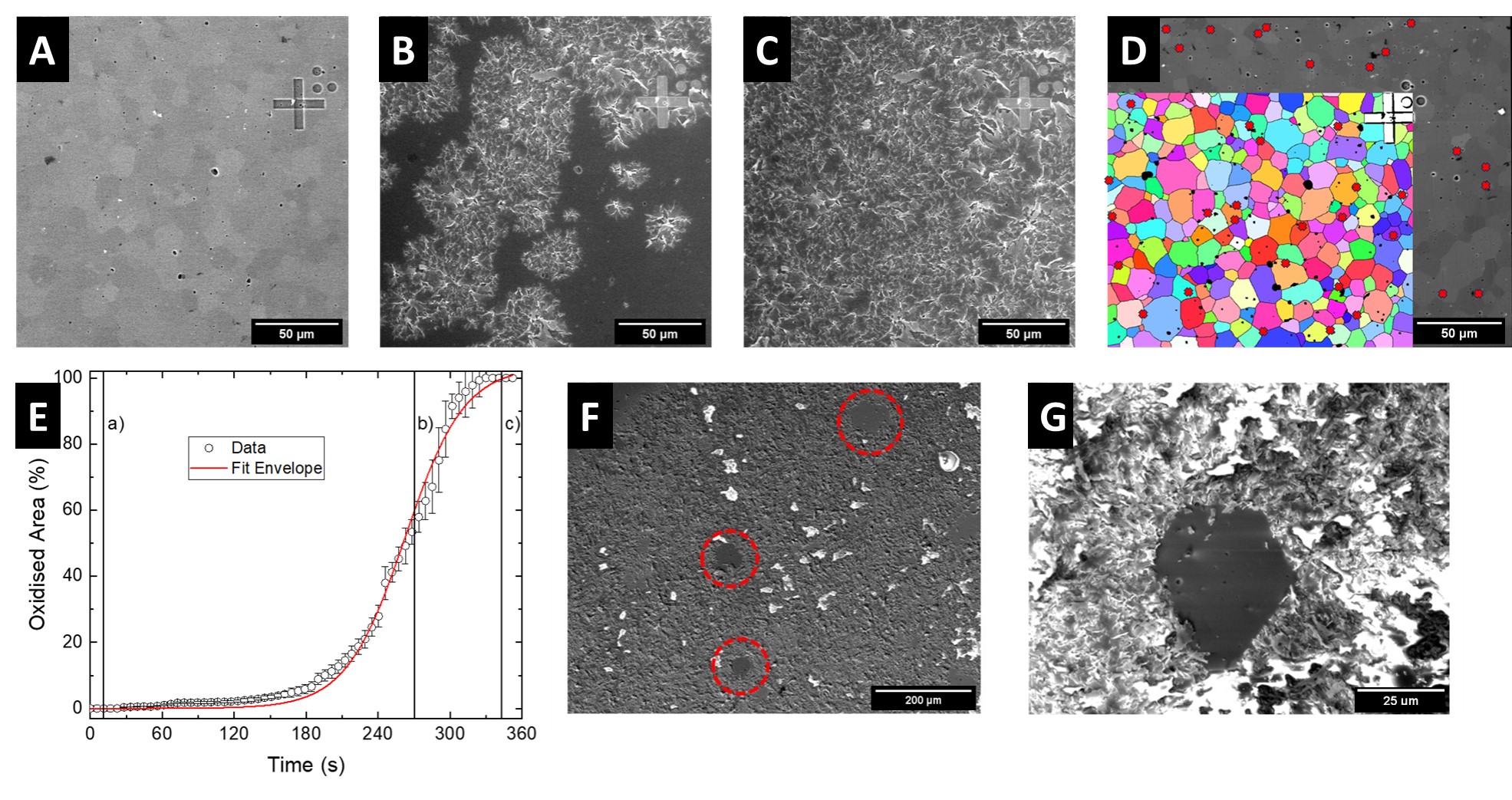}
\caption{\textbf{\textit{In-situ} HT-ESEM oxidation of polyepitaxial UO\textsubscript{2} thin films}.(A) SEM image of marked area of polyepitaxial UO\textsubscript{2} thin film before oxidation, (B) example of an SEM image taken during partial oxidation to U\textsubscript{3}O\textsubscript{8} and (C) SEM image taken after full oxidation to U\textsubscript{3}O\textsubscript{8}. The cracking of the thin films is attributed to the volume expansion associated with this phase transition. (D) an overlap of EBSD map and SEM image. Red markers indicates locations where cracking related to formation of U\textsubscript{3}O\textsubscript{8} started. From a total of 28 points where the cracking started, 21 points were located on the grain boundaries (75\%). This indicates that not only the crystallographic orientation of the grain has to be take into consideration, but also the density of the grain boundaries. (E) shows kinetics extracted from the SEM images based on the ratio of the fractured area to the entire area. With reference to (A-C), data are represented by open circles, and the fit is represented by a solid red line. Typical nucleation and growth mechanism is observed. (F) SEM image taken after sample oxidation showing some grains that did not lose integrity (G) shows higher magnification of a U\textsubscript{3}O\textsubscript{8} grain mirroring the initial UO\textsubscript{2} grain.}\label{Fig3}
\end{figure}

\subsection{In-situ electron observation of oxidation}

To further explore the potential of a polyepitaxial system, a UO\textsubscript{2} sample was oxidised in-situ in a High Temperature Environmental Scanning Electron Microscope (HT-ESEM). Prior to oxidation an area of sample was marked and mapped using EBSD. Selected images taken during this in-situ oxidation experiment are shown in Fig. \ref{Fig3}A-C. Clear cracking of the surface is observed as a result of the 36\%{} volume expansion \citep{mceachern1998review, rousseau2006detailed, quemard2009origin, elorrieta2016detailed} associated with transition to U\textsubscript{3}O\textsubscript{8}. The kinetics extracted from this process as the percentage of the damaged area to the entire area are provided in Fig. \ref{Fig3}E. The characteristic nucleation and grow character is also independently
observed in this experiment.

The ignition centers, observed as cracks related to oxidation, were extracted and are indicated by red markers on the overlay of SEM and EBSD maps in Fig. \ref{Fig3}D. In 75\%{} of cases the oxidation-related cracking was observed at the grain boundaries, which is expected  and has been previously demonstrated that oxidation preferentially occurs along the grain boundaries, and then advances from boundary to the centre of the grains \citep{mceachern1998review, liu2021grain}. This shows that faster oxidation rate correlate to regions of UO\textsubscript{2} with higher grain boundary density. Post-oxidation examination of the sample revealed U\textsubscript{3}O\textsubscript{8} grains within the polyepitaxial sample that did not disintegrate, Fig. \ref{Fig3}F-G. This would suggest a unique crystallographic relationship between between UO\textsubscript{2} and U\textsubscript{3}O\textsubscript{8} along the UO\textsubscript{2} [001] direction.

\section{Discussion}\label{sec12}

The combined use of XRD, SEM and EBSD on polyepitaxial UO\textsubscript{2} deposited onto specially engineered YSZ substrates has demonstrated that it is possible to mimic the grain structure of the nuclear fuel. This was achieved through
different heat annealing treatments of the substrate producing similar variations in
the grain size morphology of the thin film to that observed in pristine and irradiated sintered pellets of UO\textsubscript{2}. The polycrystalline character of the system, was confirmed by XRD. EBSD results showed that the grain size could be controlled within 10-50 $\mu$m without needing to expose UO\textsubscript{2} to high temperatures. Althought, other have discussed grain-to-grain epitaxy \citep{hwang1988imaging, bhat2020grain} or domain-matching epitaxy \citep{estandia2020domain}, non has realised the power of this method to controllably engineer grain structure to study material properties in the way presented here. This new high-quality system with constant lateral grain structure, and ability to acquire desired grain boundary density by controlling the average grain size, could open new areas of further investigation.     

Here we have used it to look at the UO\textsubscript{2} $\Rightarrow$ U\textsubscript{3}O\textsubscript{8} oxidation transformation. Data collected using in-situ XRD and SEM approaches has shown good agreement with the characteristic kinetics of the transition \citep{mceachern1998review}, and the importance of the grain boundary density \citep{campbell1989oxidation, taylor1980x, guenther1984dry, campbell1989oxidation2, cantwell2020grain}. The SEM data confirms recent simulation studies, were it has been shown that oxidation predominantly takes place along the grain boundaries before advancing from the boundaries towards the center of grains \citep{liu2021grain}. Our data shows that the [001] orientation is most readily oxidised, in contradiction to previous studies which report a [111]$>$[110]$>$[001] order of oxidation rate \citep{allen1986formation, allen1987oxidation}. The sample system demonstrated here can be used for further studies to confirm faster oxidation in fuel with finer grain due to the faster oxygen diffusion in the grain boundary \citep{liu2021grain, cantwell2020grain, mceachern1998review}.

Notably, although most grains transforming from UO\textsubscript{2} to U\textsubscript{3}O\textsubscript{8} experienced a high degree of disintegration due to the extreme volume expansion (36\%), a small number of grains remained laterally unaffected, and appeared to expand along the surface normal. This suggests a possible crystallographic relationship between UO\textsubscript{2} and U\textsubscript{3}O\textsubscript{8} that needs further study.

\section{Methods}\label{sec11}

\subsection{Reactive DC magnetron sputtering}\label{subsec2}

Polyepitaxial thin film samples were fabricated using reactive DC magnetron sputtering system at the FaRMS (Facility for Radioactive Materials Surfaces) at the University of Bristol \citep{springell2023review}. The system operates at a base pressure of 1 x 10\textsuperscript{-10} mbar, with 5.5N argon used as the main sputtering gas at a pressure of 7.3 x 10\textsuperscript{-3} mbar. To grow polyepitaxial UO\textsubscript{2} a depleted uranium target was used, and a partial pressure of 2 x 10\textsuperscript{-5} mbar of O\textsubscript{2}. Substrates were held at elevated temperature of 650\textdegree{}C to assist crystalline growth of UO\textsubscript{2}. With calibrated deposition rate, the growth time was controlled to achieve roughly a 100 nm thick layer of UO\textsubscript{2} \citep{springell2023review}.

\subsection{Substrate preparation}\label{subsec2}

To investigate the effect of grain structure on UO$_{2}$ oxidation, polyepitaxial UO$_{2}$ thin films were grown on polycrystalline ceramic of yttrium stabilized zirconia (Y\textsubscript{2}O\textsubscript{3})\textsubscript{0.08}(ZrO\textsubscript{3})\textsubscript{0.92} substrates obtained commercially from MTI Corporation, with dimensions 10 $\times$ 10 $\times$ 0.5 mm, polished to \textless 100 \AA{} root mean square (rms) roughness.

The structure of uranium dioxide is cubic fluorite, with space group of UO\textsubscript{2} is Fm$\overline{3}$m, No. 225, with a lattice constant at room temperature 5.47 \AA \citep{mceachern1998review, barrett1982preparation, swanson1955standard, desgranges2009neutron}. A good match between the thin film and the substrate helps to reduce lattice strain and structural defects.   

In the work of Bao \textit{et al}. \cite{bao2013antiferromagnetism} a match between [001] oriented thin film of UO\textsubscript{2} and the [001]-oriented substrates of calcium fluoride (CaF\textsubscript{2}) and lanthanum aluminate (LAO) was reported. The yttria-stabilized zirconia (YSZ) substrate to grow a [001] single crystal of UO\textsubscript{2} was reported by Strehle \textit{et al}. \cite{strehle2012characterization}. Later all three orientations [111], [110] and [001] of YSZ were used by Rennie \textit{et al}. \cite{rennie2018role} to produce single crystals of UO\textsubscript{2} for corrosion studies. In this project, YSZ was selected due to its suitability for surface preparation, given its ease of mechanical polishing and relatively straightforward induction of grain growth.

Lateral grain sizes of the YSZ substrates was increased by annealing them in a tube furnace. This process was carried out in air atmosphere, at 1600\textdegree{}C. The heating ramp rate was set to 15\textdegree{}C/min, and the same rate was used for cooling down. During heat treatment the roughness of the material increased due to changes in the grain size and shape, to obtain a smooth and clean surface, polishing and cleaning processes were applied.

To assist with the surface preparation process, annealed substrates were encapsulated in ClaroCit resin (Struers). The first stage of polishing process for one side polished YSZ was P1200, followed by P2500 and P4000 silicon carbide (SiC) abrasive papers (Buehler). Time needed for each step was between 3 minutes to 20 minutes, based on the surface quality assessment performed using an optical microscope. The polishers used during this stage were Buehler Metaserv, operated at speed of 600 rpm, with water flow. The ECOMET polishers were used in next stage. Diamond pastes with 3um, 1 um, 0.25 um and 0.1 um sizes were used, together with MetaDi fluid lubricant (Buehler). The three polishers with higher grades, use MasterTex polishing cloths fitted to a wheel, while the last one was fitted with MicroCloth. The last step of the polishing process involves use of a Vibro-Polisher (with 50 nm of Colloidal Silica), at medium vibration speed for 24 hours.

The initial cleaning stage involved ultrasonicating the substrate in a 500 ml beaker filled with 250 ml acetone for 1 hour to remove larger resin contaminants from the sample. Once this step was completed, the sample was rinsed with acetone and underwent an additional 45 minutes of ultrasonication in clean acetone. Subsequently, to further eliminate organic contaminants, the substrates were immersed in ethanol and subjected to ultrasonic treatment for one hour, then rinsed with ethanol. The same procedures were then repeated using isopropanol. Following the cleaning process with these three solvents, the surface was dried using a flow of warm, clean air to prevent residue buildup.

\subsection{In-situ X-ray analysis}\label{subsec2}

X-ray diffraction (XRD) measurements were performed using a Philips X'Pert PRO MPD diffractometer with a Cu-K$\alpha$ source, configured for Bragg–Brentano diffraction geometry, with specialized mirror designed to reflect Cu-K$\beta$ away from the detector. The footprint width of the X-ray beam on the sample surface was 10 mm, and the incident programmable divergence slit, located 240 mm away from the sample, was set to 0.5\textdegree{}. The step size was 0.05\textdegree{} and counting time per step was 2s, with sample being held stationary. The in-situ oxidation experiment was performed utilizing the Anton Paar HTK 1200 high temperature environment chamber, each time at a constant oxygen pressure of 200 mbar. Lattice parameters were extracted using a least-squares fitting routine of Pseudo-Voigt peak profiles.

\subsection{SEM characterization}\label{subsec2}

The initial sample characterisation, including EBSD, measurement were conducted at an acceleration voltage of 30 keV and aperture of 120 $\mu$m using Zeiss Sigma High Definition Variable Pressure Field Emission SEM with EDAX Electron Backscatter Diffraction detector. To analyse the data OIM Analysis\textsuperscript{TM} (EDAX) software was used. The same microscope was used to characterise sample surface post-oxidation. 

The in-situ SEM oxidation experiment was performed using a HT-ESEM at the Institut de Chimie Separative de Marcoule (ICSM) in Marcoule, France, model: FEI Quanta 200 FEG ESEM, Thermo Fisher Scientific, Massachusetts, USA. The insertion was performed in air. Samples were heated under an atmosphere of 3.5 mbar of oxygen. Once the signs of the oxidation were visible, temperature was held constant. The sample regions underwent continuous monitoring throughout the entire duration of the experiment, with images being recorded at intervals of 3 to 5 seconds. All pictures were recorded with the same magnification of x250, corresponding to an area of 512 x 470 $\mu$m$^2$.

Fiji ImageJ \citep{schneider2012nih} software, with in-built functions, was used to analyse the surface changes during the sample treatment. The sample surface cracking due to oxidation and further expansion of the crack propagation was monitored by measuring the percentage of the damaged surface to the entire ROI. The background of each photo was generated using a Gaussian-blurred version of the image and then subtracting it from the original. To enhance the contrast between the oxidised and the pristine areas of the sample, a series of find edges and smooth functions were used. In the next step Gaussian-blurred or Medina filters were applied.

\backmatter

{\section*{Data Availability}

The data will be available in the University's Data Repository in a form suitable for long-term retention and wider publication. Available from the corresponding author upon request.}

{\section*{Acknowledgements}

This research was supported by the Bristol Centre for Functional Nanomaterials, Centre for Doctoral Training, the University of Bristol Facility for Radioactive Materials Surfaces (FaRMS), funded by the UK Engineering and Physical Sciences Research Council [EP/V035495/1], and the TRANSCEND consortium on nuclear waste and decommissioning, also funded by the UK Engineering and Physical Sciences Research Council [EP/S01019X/1].}

{\section*{Author Contributions}

JW - Conceptualisation, substrate preparation, sample fabrication, EBSD data collection and analysis, in-situ XRD experiment and data analysis, SEM data collection and analysis, HT-ESEM data analysis, figures, original draft preparation
JS - substrate preparation, sample preparation, EBSD data collection and analysis 
RP - In-situ HT-ESEM data collection and analysis
JL - sample preparation, XRD data collection,
RS - Conceptualisation, data analysis, supervision, manuscript review and editing
All authors discussed and contributed to the writing of the paper.}

{\section*{Competing Interests}

The authors declare no competing interests.}

\bigskip
\begin{flushleft}%


\bigskip\noindent


\end{flushleft}


\bibliography{sn-bibliography}

\end{document}